\documentclass[a4paper,11pt,nohyper]{JHEP3}
\usepackage{amsmath,latexsym,amssymb,slashed}
\usepackage{graphicx}
\usepackage[latin1]{inputenc}

\newcommand\beal{\begin{align}}
\newcommand\nn{\nonumber}

\newcommand{\eq}[1]{\begin{equation}#1\end{equation}}
\newcommand{\spl}[1]{\begin{split}#1\end{split}}

\newcommand{\arXividhepth}[1]{\href{http://arxiv.org/abs/#1}arXiv:{\tt #1} [hep-th]}

\newcommand{\wj}{\widetilde{J}}
\newcommand{\reo}{\mathrm{Re}~\!\omega}
\newcommand{\imo}{\mathrm{Im}~\!\omega}
\newcommand{\ads}{AdS_4}

\newcommand{\boxedeq}[1]{
\begin{equation}
\fbox{
\rule[0.7cm]{0pt}{0pt}
$#1$
\rule[-0.45cm]{0pt}{0pt}
}
\end{equation}
}

\def\d{\text{d}}

\def\slashchar#1{\setbox0=\hbox{$#1$}           
\dimen0=\wd0                                 
\setbox1=\hbox{/} \dimen1=\wd1               
\ifdim\dimen0>\dimen1                        
\rlap{\hbox to \dimen0{\hfil/\hfil}}      
#1                                        
\else                                        
\rlap{\hbox to \dimen1{\hfil$#1$\hfil}}   
/                                         
\fi}

\title{New supersymmetric $\ads$ type II vacua}

\author{Dieter L\"{u}st${}^{\diamondsuit\clubsuit}$ and 
 Dimitrios Tsimpis${}^{\clubsuit}$  \\

\begin{itemize}
  
\item  Max-Planck-Institut f\"ur Physik\\
F\"ohringer Ring 6, 80805 M\"unchen, Germany
  
\item  Arnold-Sommerfeld-Center for Theoretical Physics\\
Department f\"ur Physik, Ludwig-Maximilians-Universit\"at M\"unchen\\
Theresienstra\ss e 37, 80333 M\"unchen, Germany
  \end{itemize}

\bigskip
 E-mail:
 \email{dieter.luest@lmu.de} \& \email{luest@mppmu.mpg.de}, \email{dimitrios.tsimpis@lmu.de}}

\abstract{Building on our recent results on dynamic $SU(3)\times SU(3)$ structures
 we present a set of  sufficient conditions 
for supersymmetric $\ads\times_w\mathcal{M}_6$ backgrounds of 
type IIA/IIB supergravity. These conditions
 ensure that the background  solves, besides the supersymmetry equations, 
all the equations of motion of type II supergravity.
  The conditions 
 state   
that the internal manifold is  locally a codimension-one foliation such that the five dimensional leaves 
 admit a Sasaki-Einstein structure. In type IIA the 
supersymmetry is $\mathcal{N}=2$, and the six-dimensional internal space 
is locally an $S^2$ bundle over a four-dimensional 
K\"{a}hler-Einstein base; in IIB the  internal space is the direct product of a circle and 
a five-dimensional squashed Sasaki-Einstein manifold.  
Given any five-dimensional Sasaki-Einstein manifold, 
we construct the corresponding families of type IIA/IIB vacua. 
The precise profiles of all the fields are  
determined at the solution  
and depend on whether one is in IIA or in IIB. In particular the background does not contain any sources, 
all fluxes (including the Romans mass in IIA) are generally non-zero, 
and the dilaton and warp factor are non-constant. }

\preprint{MPP-2009-77\\LMU-ASC 27/09
}

\begin{document}
\setcounter{footnote}{0}
\renewcommand{\thefootnote}{\arabic{footnote}}
\setcounter{section}{0}
\section{Introduction}
\label{introduction}

In the absence of sources and higher-order derivative corrections, 
supersymmetric backgrounds of type II supergravity 
of the form $\mathbb{R}^{1,3}\times \mathcal{M}_6$ require the internal manifold 
$\mathcal{M}_6$ to be Calabi-Yau. Moreover all  background  fluxes must be set to 
zero, resulting in  $\mathcal{N}=2$ supersymmetry in four dimensions. 
Turning on the background fluxes while preserving maximal symmetry in the four non-compact dimensions  
forces the background to be of the form of a warped product $AdS_4\times_w \mathcal{M}_6$, where
 $\mathcal{M}_6$ is no longer a Calabi-Yau.

The departure from the Calabi-Yau condition in the presence of fluxes 
can be elegantly described 
by reformulating the supersymmetry conditions in the 
framework of generalized geometry \cite{hitchin, gualtieri}. This leads to 
the statement that  $\mathcal{M}_6$ must possess a pair of compatible
 pure spinors obeying certain differential conditions \cite{gran}. 
However, the conditions from generalized geometry 
in the presence of fluxes 
are necessary but not sufficient, and therefore are not quite on the same footing  
as the Calabi-Yau condition in the absence of fluxes: Even if one 
allows for the presence of supersymmetric sources, the (generalized) Bianchi identities of all form fields 
must be imposed {in addition} in order to ensure that
 all the  equations of motion are solved \cite{kt}.

Necessary and sufficient conditions for supersymmetric solutions have been established in the case of
  (massive) IIA backgrounds with  constant 
dilaton and warp factor  in \cite{lt}. 
More recently in \cite{ltb} we presented what we called the `scalar ansatz', 
an ansatz which solves the supersymmetry conditions of type II supergravity for backgrounds where the 
internal manifold possesses  $SU(3)\times SU(3)$ structure. We were, however, unable to present 
solutions of the full set of supergravity equations of motion in the absence of sources.

In the present paper, building on the results of \cite{ltb}, we present a set of  sufficient conditions for 
supersymmetric solutions of the full set of equations of motion of type IIA/IIB supergravity in the absence of 
sources. The conditions can be concisely stated as follows. Let the background be of the form of a warped product 
 $AdS_4\times_w \mathcal{M}_6$, and let  the internal manifold 
 $\mathcal{M}_6$ be {locally} (but not necessarily globally) expressed as a codimension-one foliation:
\eq{\label{1.0}
ds^2(\mathcal{M}_6)=\d t^2+ds_t^2(\mathcal{M}_5)
~,}
where the metric of the five-dimensional leaves depends in general on the coordinate $t$. Let us moreover 
assume that on $\mathcal{M}_5$ there are three real two-forms $\alpha$, $\beta$, $\gamma$ and a real 
one-form $u$ such that:
\boxedeq{\spl{\label{intsb6}
\iota_{u}\alpha&=\iota_{u}\beta=\iota_{u}\gamma=0\\
\alpha\wedge\beta&=\beta\wedge\gamma=\gamma\wedge\alpha=0\\
\alpha\wedge\alpha&=\beta\wedge\beta=\gamma\wedge\gamma\neq0~
}}
and
\boxedeq{\label{intsb2}
\d u=-2\gamma~;~~~~~
\d(\alpha+i\beta)=-3iu\wedge(\alpha+i\beta)~;~~~~~
\d\gamma=0
~.}
%
It then follows that $AdS_4\times_w \mathcal{M}_6$, where the 
six-dimensional internal manifold is given by  (\ref{1.0}), is a supersymmetric pure-flux background (i.e. it does not contain any sources) 
of type IIA/IIB. 
As we show in appendix \ref{se}, 
the conditions in eqs.~(\ref{intsb6},\ref{intsb2}) 
are equivalent to the statement that $\mathcal{M}_5$ admits a Sasaki-Einstein structure.

We emphasize that, provided 
(\ref{intsb6},\ref{intsb2}) hold, there is no obstruction 
to specifying appropriate profiles for all 
supergravity fields so that 
the background $\ads\times_w\mathcal{M}_6$ solves all the equations of motion of type II supergravity, 
not only the supersymmetry conditions. Hence eqs.~(\ref{intsb6},\ref{intsb2}) may be viewed as replacing
 the Calabi-Yau condition in the presence of fluxes. 
The precise profiles of all the fields are  
determined at the solution, as we explain in detail in the main text, 
and depend on whether one is in IIA or in IIB. In particular, all fluxes 
(including the Romans mass in IIA) are generally non-zero, 
and the dilaton and warp factor are non-constant.

The proof of the above statements relies on our results in \cite{ltb}. In the present paper we construct 
backgrounds of the form (\ref{1.0},\ref{intsb6},\ref{intsb2}) which as we show  
 fulfill all the conditions of the scalar ansatz of \cite{ltb}, thereby solving 
the supersymmetry conditions of type II supergravity. 
Moreover we show that  all the Bianchi identities are satisfied without the need to 
introduce any source terms. Thanks to the integrability theorem mentioned earlier, this then implies that 
all the remaining equations of motion are satisfied. 

The solutions presented here are expressed in terms of the real forms $\alpha$, $\beta$, $\gamma$, $u$ mentioned 
 above, specifying a Sasaki-Einstein $SU(2)$ structure on $\mathcal{M}_5$.  On the other hand, 
the scalar ansatz of \cite{ltb} is expressed in terms of a local $SU(2)$ structure given by the triplet 
$(K,\omega,\widetilde{J})$, as reviewed in the main text, specifying 
an $SU(3)\times SU(3)$ structure on $\mathcal{M}_6$. The translation between the two
descriptions  is established 
by expressing 
the data of the  local $SU(2)$ structure in terms of $\alpha$, $\beta$, $\gamma$, $u$. 
The precise dictionary is given in eqs.~(\ref{sb1},\ref{sb1def}) below for IIA and eq.~(\ref{s1}) for IIB.

 In particular 
the metric of $\mathcal{M}_6$ can be read off of the local $SU(2)$ structure $(K,\omega,\widetilde{J})$, as explained in \cite{ltb}. The metric on the five-dimensional leaves of $\mathcal{M}_6$ picked by the supersymmetric 
solution is not Sasaki-Einstein, since it will not in general be the same as the 
metric compatible with the Sasaki-Einstein structure. 
As explained in appendix \ref{se}, the metric of the supersymmetric background 
is related to the Sasaki-Einstein metric through warping and squashing; the precise 
relation is given in the main text. In type IIA the total six-dimensional internal space 
is locally an $S^2$  bundle over a four-dimensional 
K\"{a}hler-Einstein manifold, whereas in IIB it is the direct product of a circle and 
a five-dimensional squashed Sasaki-Einstein manifold.

Given any five-dimensional (regular or not) 
Sasaki-Einstein manifold (explicit 
examples thereof are the round $S^5$, the homogeneous 
metric on $T^{1,1}$, and the infinite $Y^{p,q}$ series \cite{gma}), 
we construct the corresponding families of pure-flux vacua of type II supergravity.\footnote{Note, however, 
that only in the case of a regular five-dimensional 
Sasaki-Einstein manifold can the corresponding type IIA solution 
 have a global extension.}   
On the other hand, under the assumption of regularity, there is a  correspondence between 
 five-dimensional Sasaki-Einstein metrics and four-dimensional K\"{a}hler-Einstein manifolds of 
positive curvature \cite{fk}. Hence  for 
every four-dimensional K\"{a}hler-Einstein manifold of 
positive curvature there is a corresponding family of 
vacua of type II supergravity.

Recently, the case where $\mathcal{M}_6$ is a certain circle reduction of $M^{1,1,1}$
  was analyzed by Petrini and Zaffaroni in \cite{pz}, and belongs to the families 
of vacua presented here. 
As in \cite{pz}, the examples of section \ref{deformed} can be viewed as `massive deformations' of 
those of section \ref{sourcel}, and are given in terms of a system of two coupled first-order 
differential equations for two unknowns. 
Massive deformations of general $\ads\times\mathcal{M}_6$ backgrounds 
 were recently constructed in \cite{tgb} to first 
order in a perturbative expansion in the Romans mass.

The outline of the remainder of the paper is as follows: 
After a brief review of the type IIA scalar ansatz of \cite{ltb} 
 in the next section, we start  
with the case of $\mathcal{N}=2$ IIA compactifications with zero Romans mass in section \ref{sourcel}. 
This is subsequently 
generalized to $\mathcal{N}=2$ massive solutions with dynamic $SU(3)\times SU(3)$ structure in section \ref{deformed}. 
In section 
\ref{summary} we review the 
scalar ansatz in the case of IIB and give the solution in closed form. 
On the IIB side we only treat the static $SU(2)$ case, although we expect the dynamic $SU(3)\times SU(3)$ case to 
be a straightforward generalization thereof. 
The appendix \ref{se} contains some useful facts about five-dimensional 
Sasaki-Einstein metrics, and in particular explains the relation between
 the metric of the five-dimensional leaves of (\ref{1.0}) and 
the Sasaki-Einstein metric associated with the structure (\ref{intsb6}, \ref{intsb2}). 
Our conclusions are contained in section \ref{conclusions}. 
We have collected some useful formul\ae{} in the appendix.

\section{The IIA side}

In the following subsection we start by reviewing the scalar ansatz of \cite{ltb}, specialized to the case of IIA. For 
more details the reader may consult that reference. Then in  subsection \ref{deformed} we present the 
$\mathcal{N}=2$ solutions, after a brief discussion of the zero Romans mass limit in section \ref{sourcel}.

\subsection{Review of the scalar ansatz}\label{review}

The ten-dimensional spacetime metric (in the string frame) 
is given by:
\eq{\label{1m}
ds^2=e^{2A}ds^2(\ads)+ds^2(\mathcal{M}_6)
~,}
where $A$ is the warp factor. 
The internal
 six-dimensional manifold $\mathcal{M}_6$ is characterized by a local $SU(2)$ structure determined 
 by the triplet $(\omega,\widetilde{J}, K)$, where 
$\omega$ is a complex two-form, $\widetilde{J}$ is a real two-form, and 
$K$ is a complex one-form. These forms satisfy the following algebraic compatibility conditions:
\eq{\spl{\label{su2alg}
\wj\wedge\omega&=0\\
\wj\wedge\wj&=\reo\wedge\reo=\imo\wedge\imo\neq0\\
\iota_K\wj&=\iota_K\reo=\iota_K\imo=0
~.}}
Moreover, as explained in \cite{ltb}, 
associated with this local $SU(2)$ structure there are two {\it global} $SU(3)$ structures 
$(J^{(i)},\Omega^{(i)})$, $i=1,2$, given by:
\eq{\spl{
J^{(1)}=\frac{i}{2}K\wedge K^*+ \widetilde{J}~&; ~~~~~J^{(2)}=\frac{i}{2}K\wedge K^*- \widetilde{J}\\
\Omega^{(1)}=-i{\omega}\wedge K~&;~~~~~\Omega^{(2)}=i{\omega}^*\wedge K
~,}}
where we have normalized $|K|^2=2$.

The scalar ansatz introduced in \cite{ltb} is an ansatz which solves the supersymmetry conditions 
of type II supergravity for backgrounds where the internal six-dimensional manifold possesses
 $SU(3)\times SU(3)$ structure. According to the ansatz one truncates the components of all the form 
fluxes to those which are singlets with respect to the local $SU(2)$ structure in $(\ref{su2alg})$. More specifically, 
the NSNS three-form is given by
\eq{\label{nsfluxesiia}
H= \frac{1}{24}\left(
h_1\omega^* +h_2~\!\omega+2h_3\widetilde{J} 
\right)\wedge K
+\mathrm{c.c.}
~,}
while the RR fluxes are given by:
\beal\label{b16}
e^{\phi}F_0&=f_0\nn\\
e^{\phi}F_2&=\frac18 \left(
 f_2~\!\omega^*+ f_3\widetilde{J} +2i f_1K\wedge K^*\right)
+\mathrm{c.c.}\nn\\
e^{\phi}F_4&=
\frac{1}{16}g_1\widetilde{J}\wedge \widetilde{J}
+\frac{i}{96}\left(g_2~\!\omega^*+g_2^*~\!\omega +2g_3\widetilde{J}  
\right)\wedge K\wedge K^*\nn\\
e^{\phi}F_6&=f~\! vol_6
~,
\end{align}
where the various scalar coefficients above are given by eq.~(2.15) of \cite{ltb}.\footnote{Note that 
there was a typo in the last line of that equation in the previous versions of ref.~\cite{ltb}.} Moreover, 
the local $SU(2)$ structure is constrained to obey the differential conditions eqs.~(2.26) of \cite{ltb}. 
In addition one must impose the constraints in eqs.~(2.16,2.17) of \cite{ltb}. 

As explained in \cite{ltb}, in constructing explicit solutions the non-trivial task is to 
find manifolds admitting local $SU(2)$ structures such that they obey the differential conditions 
mentioned above. Moreover one has to worry about the Bianchi identities, which were not considered in \cite{ltb}. 
In the following subsection we present families of solutions obeying 
all the conditions of the scalar ansatz, thus solving the supersymmetry equations of IIA. 
In fact, as we will see, the supersymmetry is $\mathcal{N}=2$ in four dimensions (eight real supercharges). 
In addition we show that all the Bianchi identities are satisfied, without the need to introduce any sources. 
In other words the solutions correspond to pure-flux backgrounds.

\subsection{Undeformed solutions}\label{sourcel}

Our general solutions in the type IIA case can be viewed as families of solutions 
partly parameterized by the Romans mass. These families include as 
special cases 
the solutions with zero  Romans mass. These special cases correspond to backgrounds which can be uplifted to
 solutions of eleven-dimensional supergravity and have already appeared 
in the literature  \cite{gm}: 
the resulting seven-dimensional internal manifold is Sasaki-Einstein,  
 as follows from the properties of Freund-Rubin vacua, and 
for a global solution it must belong to the $Y^{p,q}$ series\footnote{Note however that our procedure for constructing IIA solutions can be carried out starting with any five-dimensional Sasaki-Einstein manifold, regular or not. In the latter case the SE manifold cannot be thought of as the total space of a line bundle over a globally-defined four-dimensional K\"{a}hler-Einstein manifold. The construction in our paper can nonetheless still be carried out to produce a local solution. In that case the massless, undeformed limit of the 
solution will not be a $Y^{p,q}$ reduction -- since the latter would require 
a globally-defined, four-dimensional, smooth, positive-curvature K\"{a}hler-Einstein base.}.

As explained in \cite{tga}, on the dual CFT$_3$ side the mass deformation of the supergravity background corresponds 
to the sum of the levels of the two Chern-Simons terms.
 As in \cite{pz}, the deformed solutions are presented here explicitly up to 
a coupled system of two first-order differential equations 
for two unknowns.

Before coming to the general (`deformed') solutions  in section \ref{deformed}, we review here the special (`undeformed') 
solutions with zero Romans mass. As we will see, the former correspond to 
backgrounds with dynamic $SU(3)\times SU(3)$ structure, while the latter possess strict $SU(3)$ structure. 
In both cases the solutions possess $\mathcal{N}=2$ supersymmetry.

The supersymmetry equations 
and Bianchi identities for the strict $SU(3)$ structure case are summarized in appendix \ref{iiaside}. We can  
make use of the scalar ansatz by expressing the strict $SU(3)$ structure in terms of the 
 local $SU(2)$ structure of $\mathcal{M}_6$. Furthermore, we will take the latter to be  given by:
\eq{\spl{\label{sb1}
K&=e^{A}\left(\xi u+i\d t\right)\\
\frac{1}{3}\widetilde{J}&=e^{2A}\left(\sin\theta~\!\alpha+\cos\theta~\!\gamma\right)\\
\frac{1}{3}\omega&=e^{2A}\left(\cos\theta~\!\alpha-\sin\theta~\!\gamma+i\beta\right)
~,}}
where $\alpha$, $\beta$, $\gamma$ are real two-forms on $\mathcal{M}_5$ and $u$ is a real one-form on $\mathcal{M}_5$ obeying 
(\ref{intsb6},\ref{intsb2}); we will assume that $A$, $\theta$, $\xi$ are 
all functions of $t$.

As explained in appendix \ref{se}, it follows from the above that 
the metric of the six-dimensional space $\d s^2(\mathcal{M}_6)$ is of the form 
of a codimension-one foliation:
\eq{\label{sb5}
ds^2(\mathcal{M}_6)=e^{2A}\left(ds_t^2(\mathcal{M}_5)+dt^2\right)
~,}
where the metric of the five-dimensional leaves is given locally by:
\eq{\label{3.3}
ds_t^2(\mathcal{M}_5)=3ds^2_{KE}+\xi^2~\!u\otimes u~.
}
For general $\xi$ this is a squashed Sasaki-Einstein metric. The metric is locally of the form 
of a $U(1)$ fibration with connection field-strength given by $\d u=-2\gamma$. 
The four dimensional base over which $u$ is fibered is 
locally K\"{a}hler-Einstein with metric $ds^2_{KE}$.

{}Furthermore, we will assume that all fluxes are zero except for $F_2$, $F_6$, which 
are given by (\ref{nzf}). 
Plugging the ansatz  into the first of (\ref{torsiia}) and 
setting $\mathrm{Im}W=1$ for simplicity\footnote{Remember 
that $W$ is the inverse $AdS$ radius and is therefore a constant.}, 
we obtain the following three equations:
\eq{\spl{\label{4.1}
\xi&=\frac{3}{2 }\sin\theta\\
A'&=\frac{1}{2 }\sin\theta\cos\theta\\
\theta'&=1+{\cos^2\theta }
~,}}
where the prime denotes differentiation with respect to $t$. 
The first equation above determines $\xi$ in terms of the angle $\theta$, while the last two can be solved to determine 
$\theta$ and the warp factor $A$ as functions of $t$:
\eq{\spl{\label{sb55}
A&=\frac{1}{4}\log\left[1+\sin^2(\sqrt{2}~\!  ~\! t)\right]\\
\cos\theta&=\frac{\cos(\sqrt{2}~\!  ~\! t)}{\sqrt{1+\sin^2(\sqrt{2}~\!  ~\! t)}}
~.}}

The last line of (\ref{torsiia}) is automatically satisfied, by virtue of (\ref{4.1}). Moreover, the second line of 
(\ref{torsiia}) can be seen to be satisfied, again taking (\ref{4.1}) into account, provided:
\eq{\label{4.16}
\widetilde{F}={ }e^{-4A}\left(\frac{4}{3}-\sin^2\theta\right)
(\widetilde{J}+e^{2A}\sin\theta u\wedge\d t)
~.}

As we show in appendix \ref{iiaside}, in order to have a solution to all the equations of motion we only 
need to make sure that the Bianchi identity (\ref{bi2}) is satisfied. From (\ref{nzf},\ref{4.16}), taking (\ref{4.1}) into account, we can see that $F_2$  can be written in the form:
\eq{\label{f2d}
F_2=\frac{1}{2 }~\!\d\left(e^{-2A}\cos\theta~\!u \right)
~,}
and therefore it  satisfies the Bianchi identity (\ref{bi2}).

We would like to stress that one can take $ds^2(\mathcal{M}_5)$ in (\ref{3.3}) to be the squashed 
version of any one of the (infinitely-many)  
five-dimensional Sasaki-Einstein metrics, with the squashing $\xi$ given in (\ref{4.1},\ref{sb55}). 
The resulting $ds^2(\mathcal{M}_6)$ metric in (\ref{sb5}) in particular includes as a special case the metric 
of the circle reduction of the 
$M^{1,1,1}$ manifold recently analyzed in \cite{pz}. 
Note however, 
 that the case of $\mathbb{CP}^3$ is not of the form (\ref{sb5}): Although $\mathbb{CP}^3$ can indeed be thought 
of as a codimension-one foliation with leaves given by  $T^{1,1}$, the metric is 
not of the form (\ref{3.3}).

Moreover one can show that the foliations (\ref{sb5}) are smooth. 
To see this note that potential singularities 
arise at the zeros of $\xi(t)$,  which occur at $t_0=n\pi/\sqrt{2} ,~n\in\mathbb{Z}$. 
Let us set $u=d\psi+\mathcal{A}$, where $\psi$ is the coordinate of the $U(1)$ fiber and $\mathcal{A}$, such that 
$\d\mathcal{A}=-2\gamma$,  is the $U(1)$ connection. 
Near $t_0$ the metric takes the form:
\eq{\label{sb5appr}
ds^2(\mathcal{M}_6)\approx e^{2A(t_0)}\left(dt^2+9(t-t_0)^2(d\psi+\mathcal{A})^2+3ds^2_{KE}\right)
~,}
where we have taken (\ref{4.1},\ref{sb55}) into account. Assuming the 
warp factor does not blow up at $t_0$, this will be free of  
singularities provided we take $\psi$ to have period $2\pi/3$.

The above argument also shows that if we take $t$ to have the 
range $0\leq t\leq\pi/\sqrt{2} $,  (\ref{sb5}) can be thought of locally as an $S^2$ fibration over 
the four-dimensional K\"{a}hler-Einstein base. Indeed, fixing the point on the four-dimensional 
base, the $(\psi,t)$ fiber is a circle parametrized by $\psi$ which is fibered over $t$. Moreover  
the circle smoothly shrinks to 
zero size at the endpoints of the interval $t=0,\pi/\sqrt{2} $, 
showing that the $(\psi,t)$ fiber has the topology of $S^2$.

Finally let us remark that 
 the solutions presented here possess $\mathcal{N}=2$ supersymmetry. 
Indeed this is a consequence of the fact that, thanks to (\ref{f2d}), the fluxes do 
not depend on the two-forms $\alpha$, $\beta$. Moreover, as explained in appendix \ref{se}, the metric 
is invariant under general orthogonal rotations of the triplet $\alpha$, $\beta$, $\gamma$. 
Consequently all fields are invariant 
under $SO(2)$ rotations in the $(\alpha, \beta)$ plane. (Note that these rotations would have to 
be $t$-independent  for them to leave eqs.~(\ref{intsb2}) invariant.)


\subsection{Mass-deformed solutions}\label{deformed}

We are now ready to generalize the solutions of the previous subsection 
to include non-zero Romans mass and dynamic $SU(3)\times SU(3)$ structure.

We will take the  local $SU(2)$ structure of $\mathcal{M}_6$  to be  given by:
\eq{\spl{\label{sb1def}
K&=e^{B}\left(\d t-i\xi u\right)\\
\frac{W}{3}\widetilde{J}&=e^{2C}\left\{\sin\theta~\!(\cos\zeta\alpha-\sin\zeta\beta)+\cos\theta~\!\gamma\right\}\\
\frac{W}{3}\omega&=e^{2C}\left\{\cos\theta~\!(\cos\zeta\alpha-\sin\zeta\beta)-\sin\theta~\!\gamma
+i(\cos\zeta\beta+\sin\zeta\alpha)\right\}
~,}}
where we have set $W\in\mathbb{R}$. As before 
$\alpha$, $\beta$, $\gamma$ are real two-forms on $\mathcal{M}_5$ and $u$ is a real one-form on $\mathcal{M}_5$ obeying 
(\ref{intsb6},\ref{intsb2}); we take $B$, $C$, $\theta$, $\zeta$ to be  
functions of $t$. 
Note that, up to the different warp factors, the deformed 
ansatz above is obtained from the undeformed one in (\ref{sb1}) via a $t$-dependent $SO(2)$ rotation 
in the $(\alpha,\beta)$ plane through angle $\zeta(t)$.

{}Furthermore we take the spinor ansatz corresponding to the local $SU(2)$ structure above ({\it cf.} 
eq.~(2.7) of \cite{ltb}) to be given by:
\eq{\label{spinan}
\theta_1=e^{\frac{1}{2}A}\eta_1~;~~~~~
\theta_2=-e^{\frac{1}{2}A}\left(\sin\varphi~\! \eta_2^*+ie^{i\varepsilon}\cos\varphi~\! \eta_1^*\right)
~,}
with $\varphi$, $\varepsilon$ functions of $t$. These angles will turn out to be non-constant, 
thus corresponding to a dynamic $SU(3)\times SU(3)$ structure. As before, 
the warp factor $A$ is taken to be a function of $t$. In addition the angle $\theta$ in (\ref{sb1def}) 
obeys:
\eq{\label{333}
\tan\theta=\frac{\tan\varphi}{\sin\varepsilon}
~.}

It follows from the above that 
the metric of the six-dimensional space $\d s^2(\mathcal{M}_6)$ is locally of the form 
of a codimension-one foliation:
\eq{\label{sb5def}
ds^2(\mathcal{M}_6)=e^{2B}\left(\frac{3}{W}e^{2(C-B)}ds^2_{KE}+\xi^2~\!u\otimes u+dt^2\right)
~.}
Note that we could use up the reparameterization invariance of $t$ to set either one 
of $B$ or $\xi$ to some given function of $t$. This redundancy will prove useful in the following. 
{}Furthermore the fluxes are given by (\ref{b16}), where:
\eq{\spl{\label{315}
f&=-3 We^{-A} \cos \varepsilon \cos \varphi\\
f_0&=-W e^{-A}\left(\cos \varphi  \sin\varepsilon+\csc \varepsilon \sin \varphi \tan \varphi\right)\\
f_1&=-\cos \varepsilon \cos \varphi \left(We^{-A}+4e^{-B}A' \cot\varphi \sin \varepsilon \right)\\
f_2&=-8 e^{-B}A' \cos \varepsilon \cos\varphi\\
g_1&=-8\left( \cos \varphi \sin
   \varepsilon   +\sin \varphi  \csc \varepsilon \tan
   \varphi\right)
\left(W e^{-A}-4e^{-B}A' \cot\varphi \sin \varepsilon 
\right)\\
g_2&=48 \sin
   \varphi\left(We^{-A} +e^{-B}A' \cot \varphi \sin
   \varepsilon
\right)\\
h_1&=-6 \sin^2 \varphi \cot
   \varepsilon\left(
We^{-A} -2
   e^{-B}A'\sin \varepsilon  \cot
   \varphi
\right)\\
\frac{h_1}{h_3}&=\frac{h_2}{h_3}=\frac{f_2}{f_3}=\frac{g_2}{g_3}=-\tan\theta
~.
}}

With  the above equations, it is straightforward to verify that all 
conditions of the scalar ansatz of \cite{ltb} are satisfied, provided the following 
 equations hold:
\eq{\spl{\label{syst}
e^{4A}&=\frac{1}{\cos^2\theta}\tan\varepsilon\\
e^{B-A}&=-\frac{1}{2W}\cot\theta~\!(\log\tan\varepsilon)'\\
e^{\phi-3A}&=\cos\varphi\cos\varepsilon\\
\xi&=\frac{3}{2W}e^{A-B}\sin\theta\\
\zeta'&=\frac{1}{2W}e^{2(A-C)}\cos\theta\cot\varepsilon\sin^2\varphi~\!(\log\tan\varepsilon)'\\
\theta'&=\cot\theta\left(
\frac{1}{2W}e^{2(A-C)}\sin^2\theta-1
\right)(\log\tan\varepsilon)'\\
C'&=-\frac{1}{4W}e^{2(A-C)}(\sin^2\varphi+\cos^2\theta)(\log\tan\varepsilon)'
~.}}
Taking (\ref{333}) into account and using a $t$-coordinate transformation to 
fix $\theta$ to some given function of $t$,  it readily follows that the first five of 
the system of equations (\ref{syst}) 
solve for $A$, $B$, $\phi$, $\xi$, $\zeta$ in terms of $C$, $\varepsilon$. Moreover, the last two equations in 
(\ref{syst}) is a system of two coupled first-order 
differential equations for the two unknowns $C$, $\varepsilon$. This is exactly 
as in \cite{pz}. Unfortunately we will not be able 
to provide an analytical solution for this system here, but will note that 
it can be analyzed perturbatively using numerical methods \cite{pz}. 

It is a tedious but straightforward calculation to show that all the Bianchi identities 
are satisfied without further constraints. To somewhat simplify 
 the computation one may choose the `gauge' $B=A$ in order to fix the 
redundancy in the definition of the coordinate $t$. 
It is also  useful to take the formul\ae{} in appendix \ref{app} into account.

{}We can use  the same argument as 
in the undeformed case 
to show that the foliations (\ref{sb5def}) are smooth, provided 
the period of the  coordinate of the $U(1)$ fiber is chosen appropriately, 
and that the six-dimensional metric can also be thought of locally 
as an $S^2$ fibration over the four-dimensional K\"{a}hler-Einstein base. 
For example, using a $t$-coordinate transformation to 
fix $\theta$ to be the same as in the undeformed case and assuming 
the warp factors do not blow up, 
the discussion around (\ref{sb5appr}) carries over virtually 
unchanged.

Finally let us remark that, as in the undeformed case, 
 the solutions presented here possess $\mathcal{N}=2$ supersymmetry. 
This follows from the fact that, thanks to (\ref{315}), the fluxes do 
not depend on the two-forms $\alpha$, $\beta$. Consequently all fields are invariant 
under $t$-independent $SO(2)$ rotations in the $(\alpha, \beta)$ plane, 
and hence there is an  $SO(2)$-worth of $SU(3)\times SU(3)$ structures 
satisfying the supersymmetry conditions.

%

\section{The type IIB side}\label{summary}

Let us start by reviewing the scalar ansatz of \cite{ltb} for type IIB solutions, specializing 
to the case of static $SU(2)$. The ten-dimensional spacetime metric is again of the form (\ref{1m}). 
The NSNS three-form is given by
\eq{\label{nsfluxes}
H= \frac{1}{24}\left(
h_1\omega^* +h_2~\!\omega+2h_3\widetilde{J} 
\right)\wedge K
+\mathrm{c.c.}
~,}
while the RR fluxes are given by:
\eq{\spl{\label{rrfluxes}
e^{\phi}F_1&=g_1K+\mathrm{c.c.}\\
e^{\phi}F_3&=\frac{1}{24}\left(
f_1\omega^* +f_2~\!\omega+2f_3\widetilde{J} 
\right)\wedge K
+\mathrm{c.c.}\\
e^{\phi}F_5&=g_2\star_6\! K+\mathrm{c.c.}
~,}}
where the various scalar coefficients above are given by eq.~(4.1) of \cite{ltb}. Moreover, 
the static $SU(2)$ structure is constrained to obey the differential conditions eqs.~(4.3) of \cite{ltb}. 
In addition one must impose the constraints in eqs.~(4.2) of that reference.


We take the local $SU(2)$ structure to  be given by:
\eq{\spl{\label{s1}
K&=e^{A}\left(\frac{6}{5W}~\!u+i\d t\right)\\
\frac{5W^2}{6}\widetilde{J}&=e^{2A}\left(\sin\theta~\!\alpha+\cos\theta~\!\beta\right)\\
\frac{5W^2}{6}\omega&=e^{2A}\left(\cos\theta~\!\alpha-\sin\theta~\!\beta-i\gamma\right)
~,}}
where $\alpha$, $\beta$, $\gamma$ are real two-forms on $\mathcal{M}_5$ and $u$ is a real one-form on $\mathcal{M}_5$ obeying 
(\ref{intsb6},\ref{intsb2}). The corresponding six-dimensional metric reads:
\eq{\label{s5}
ds^2(\mathcal{M}_6)=e^{2A(t)}\left(ds^2(\mathcal{M}_5)+dt^2\right)
~,}
where 
\eq{\label{s5t}
ds^2(\mathcal{M}_5)=\frac{6}{5W^2}\left(ds^2_{KE}+\frac{6}{5}~\!u\otimes u\right)
~,}
and we have taken $W\in \mathbb{R}$. This is the local form of the metric of a squashed five-dimensional 
Sasaki-Einstein manifold.

The NSNS three-form is given by
\eq{\label{s7}
H= \frac{1}{2}W\mathrm{Re}\omega\wedge\d t
-\left(2A'\widetilde{J}+ce^{-4A}\mathrm{Re}\omega\right)\wedge e^{-A}\mathrm{Re}K
~,}
where  $c$ is a real constant.   
The RR fluxes are given by:
\eq{\spl{\label{s8}
e^{\phi}F_1&=-2ce^{-4A}\d t\\
e^{\phi}F_3&=-\frac{1}{2}W\widetilde{J}\wedge\d t
+\left(2A'\mathrm{Re}\omega-ce^{-4A}\widetilde{J}\right)\wedge e^{-A} \mathrm{Re}K\\
e^{\phi}F_5&=\frac{3}{2}W\widetilde{J}\wedge\widetilde{J}\wedge e^{-A}\mathrm{Re}K
~,}}
while the dilaton is related to the warp factor through:
\eq{\label{3333}
\phi=4A~.
}

It is now straightforward to verify that all the Bianchi identities are satisfied, 
provided we take:
\eq{\label{s10}
e^{4A}=\left\{
\begin{array}{cc} 
\frac{2}{\sqrt{5}}\left|\frac{c}{W}\right|\cosh\left[\sqrt{5}W(t-t_0) \right]~,& ~~~~~c\neq0\\ 
 \exp\left[\sqrt{5}W(t-t_0) \right] ~,& ~~~~~c=0
\end{array}\right.
~,}
and:
\eq{\label{s3}
\theta=\left\{
\begin{array}{cc} 
\arctan\tanh\left[\frac{\sqrt{5}}{2}W(t-t_0) \right]+\theta_0~,& ~~~~~c\neq0\\ 
 \theta_0 ~,& ~~~~~c=0
\end{array}\right.
~,}
for some constant $\theta_0$. The real constant $c$ distinguishing the two different cases 
above is the same one as in  eqs.~(\ref{s7},\ref{s8}). 
It follows that in the absence of $F_1$ flux ($c=0$) the solution is a linear dilaton background.

The ten-dimensional 
metric in the Einstein frame is of direct-product form:
\eq{
ds^2_E=ds^2(\ads)+ds^2(\mathcal{M}_5)+dt^2
~,}
as follows from (\ref{3333}). However, we suspect that this feature is an artifact 
of the static $SU(2)$ structure of the solution. We do not expect more general 
$SU(3)\times SU(3)$-structure solutions to be of direct-product form.

We  may choose to compactify the $t$-direction by a coordinate transformation.  For example,  
considering the $c\neq0$ case we can take:
\eq{
\sqrt{5}W(t-t_0)=\log\tan\frac{\chi}{4}
~,}
upon which the dilaton takes the form:
\eq{
e^{\phi}=\frac{2}{\sqrt{5}}\left|\frac{c}{W}\right|\frac{1}{\sin\frac{\chi}{2}}
~.}
Hence in the compactified description the solution  
is singular.

\section{Conclusions}\label{conclusions}

We have presented a set of sufficient conditions for the existence of supersymmetric backgrounds 
 of IIA/IIB supergravity of  the form $\ads\times_w\mathcal{M}_6$.  The conditions state that the internal six-dimensional 
manifold should be locally (but not necessarily globally) a codimension-one foliation, 
such that the five-dimensional leaves admit a  Sasaki-Einstein  structure. In type IIA the 
supersymmetry is $\mathcal{N}=2$, and the total six-dimensional internal space 
is locally an $S^2$ bundle over a four-dimensional 
K\"{a}hler-Einstein manifold; in IIB the  internal space is the direct product of a circle and 
a five-dimensional squashed Sasaki-Einstein manifold.

The solutions presented here 
 are of obvious relevance 
to the $\ads/CFT_3$ correspondence.  
Recently, the case where $\mathcal{M}_6$ is a certain circle reduction of $M^{1,1,1}$
  was analyzed by Petrini and Zaffaroni in \cite{pz}. 
As in \cite{pz}, the examples of section \ref{deformed} can be viewed as `massive deformations' of 
those of section \ref{sourcel}, and are given in terms of a system of two coupled first-order 
differential equations for two unknowns.

Massive deformations of general $\ads\times\mathcal{M}_6$ backgrounds, 
including the $\ads\times \mathbb{CP}^3$ 
background \cite{np} as a special case, were recently constructed in \cite{tgb} to first 
order in a perturbative expansion in the Romans mass. 
Both $\mathbb{CP}^3$ and the 
circle reduction of $M^{1,1,1}$ considered in \cite{pz} can be viewed as 
codimension-one foliations with five-dimensional leaves admitting Sasaki-Einstein structures. 
However, as explained in section \ref{sourcel}, only in the case of $M^{1,1,1}$ is the 
foliation of the precise form considered here; the $\ads\times\mathbb{CP}^3$ type IIA solution 
is { not} among those of section \ref{sourcel}. Instead we have a  foliation with 
$T^{1,1}$ leaves such that the total space is locally 
an $S^2$ bundle over $\mathbb{CP}^1\times \mathbb{CP}^1$.

Given any five-dimensional 
Sasaki-Einstein manifold (regular or not), we have constructed corresponding families of pure-flux vacua of type II supergravity  
with all fluxes (including the Romans mass in IIA) generally non-zero. 
Explicit examples of five-dimensional Sasaki-Einstein  spaces 
are the round $S^5$, the homogeneous 
metric on $T^{1,1}$, and the infinite $Y^{p,q}$ series \cite{gma}.\footnote{Note, however, 
that only in the case of a regular five-dimensional 
Sasaki-Einstein manifold can the corresponding type IIA solution 
 have a global extension.}   
On the other hand, under the assumption of regularity, 
 there is a  correspondence between 
five-dimensional Sasaki-Einstein metrics and four-dimensional K\"{a}hler-Einstein manifolds of 
positive curvature \cite{fk}. Hence  for 
every four-dimensional K\"{a}hler-Einstein manifold of positive curvature
 we have constructed corresponding 
families of vacua of type II supergravity.

The massless IIA vacua presented in section \ref{sourcel} 
can be uplifted to Freund-Rubin vacua of eleven-dimensional supergravity. 
The resulting seven-dimensional internal manifold must be Sasaki-Einstein,  
 as follows from the properties of Freund-Rubin vacua. 
Recall that in type IIA the internal six-dimensional 
manifold $\mathcal{M}_6$ can be viewed locally as an $S^2$  bundle over 
a four-dimensional K\"{a}hler-Einstein base. 
Hence, given any four-dimensional K\"{a}hler-Einstein manifold of 
positive curvature there is 
a corresponding seven-dimensional Sasaki-Einstein one which is a 
circle fibration over $\mathcal{M}_6$. Indeed, 
this should precisely correspond to the construction of \cite{gm}.  
This can also be seen explicitly from the type IIA reduction of the solutions 
discussed in \cite{ms2}, {\it cf.} section 5.1 therein.\footnote{ We thank D.~Martelli for bringing this 
to our attention.} In other words, the type IIA solutions presented here include the massive deformations 
of the IIA reduction of the $Y^{p,q}$ solutions discussed in \cite{ms2}.

The solutions presented here are by no means the 
most general. Even within 
the framework of the scalar ansatz, it would be interesting to try to extend our solutions by e.g. generalizing the dependence of the 
local $SU(2)$ structure on the Euler angles in eqs. (\ref{sb1def}). 
As already remarked, in the case of IIB our solutions are of the static $SU(2)$ type. 
However, we expect the generalization to dynamic $SU(3)\times SU(3)$ structure to 
be straightforward. We expect that 
for such generalized backgrounds the internal six-dimensional 
manifold would no longer possess a direct-product structure.  
We hope to report on this in the near future. 
The IIB backgrounds presented in section \ref{summary} could be Wick-rotated  
to obtain cosmological solutions with time-dependent dilaton. These may be amenable 
to analysis with conformal field theory techniques. It would be interesting to pursue this 
 further.

The results of the present paper, which 
relied on the `scalar ansatz' of \cite{ltb}, suggest that 
four-dimensional K\"{a}hler-Einstein manifolds 
play a central role in flux compactifications. 
Smooth four-dimensional K\"{a}hler-Einstein manifolds of 
positive curvature were classified in \cite{ty}: they are $\mathbb{CP}^1\times\mathbb{CP}^1$, $\mathbb{CP}^2$, 
and the del Pezzo surfaces $dP_3,\dots,dP_8$. It is intriguing that 
the latter have been shown to play a special role in recent F-theoretic constructions 
with phenomenological applications \cite{bhv}. It remains to be seen whether this mathematical structure persists
 beyond the present setup.

\appendix

\section{Five-dimensional Sasaki-Einstein manifolds}\label{se}

In this section we show that eqs.~(\ref{intsb6},\ref{intsb2}) are equivalent to the 
statement that $\mathcal{M}_5$ admits a Sasaki-Einstein structure. Moreover 
we spell out the precise relation between the Sasaki-Einstein metric and the 
metric picked by the supersymmetric background. As we will see,  the latter is  
obtained from the former by warping and squashing.

In five dimensions a Sasaki-Einstein manifold may be defined under certain 
additional mild assumptions (see for example \cite{fk}, or theorem 5.1.6 of \cite{gali})
 as one which admits a pair of Killing spinors, related by complex conjugation,  obeying:
\eq{\label{ke}
\nabla_m\eta=\pm\frac{i}{2}\Gamma_m\eta
~.}
Let us assume that $\eta_1$ is a Killing spinor obeying (\ref{ke}) with the positive sign, 
and let us define 
\eq{
\eta_2:=C\eta^*_1~.
}
It then  follows that 
$\eta_2$ obeys (\ref{ke}) with the negative sign. (For our spinor conventions the reader may consult section \ref{spinorconventions}). 
With the above normalization the metric is Einstein so that the Ricci tensor of $\mathcal{M}_5$ is given by:
\eq{
R_{mn}=4 g_{mn}
~,}
and therefore the six-dimensional cone $\mathcal{C}(\mathcal{M}_5)$ is Calabi-Yau.

It follows from (\ref{ke})  that the norm of $\eta_1$ (which is equal to the 
norm of $\eta_2$) is constant. 
We will take the norm of $\eta_{1,2}$ to be given by:
\eq{\label{44}
\eta_{1}^\dagger\eta_{1}=\eta_{2}^\dagger\eta_{2}=1
~.}
Moreover let us define a real one-form $u$:
\eq{\label{aa1}
u_m:=(\widetilde{\eta}_2\Gamma_m\eta_1)
~,}
where we used definition (\ref{wtdef}), and three real two-forms $\alpha$, $\beta$, $\gamma$:
\eq{\spl{\label{a2}
\alpha_{mn}+i\beta_{mn}&:=(\widetilde{\eta}_1\Gamma_{mn}\eta_1)=(\widetilde{\eta}_2\Gamma_{mn}\eta_2)^*\\
\gamma_{mn}&:=i(\widetilde{\eta}_2\Gamma_{mn}\eta_1)
~.}}

A little bit of Fierzing reveals that the forms defined above satisfy all the 
algebraic conditions (\ref{intsb6}). To see this, it is useful to note 
that (\ref{44},\ref{aa1},\ref{a2},\ref{fier}) imply:
\eq{
\eta_1\widetilde{\eta}_2=\frac{1}{4}\left\{
-1+u_m\Gamma^m+\frac{i}{2}\gamma_{mn}\Gamma^{mn}
\right\}
~.}
Similarly, taking the Killing spinor equation into account one can show that 
all the differential equations in (\ref{intsb2}) are satisfied.

We have thus showed 
that (\ref{intsb6},\ref{intsb2}) follow from the assumption 
that $\mathcal{M}_5$ admits a Sasaki-Einstein metric. The converse can also be seen as follows: Let 
$\mathcal{M}_5$ be an $SU(2)$-structure manifold. It follows that on $\mathcal{M}_5$ there 
is a globally-defined nowhere-vanishing spinor $\eta_1$ with associated 
$SU(2)$ structure (\ref{intsb6}). If $\mathcal{M}_5$ does not 
admit a Sasaki-Einstein  metric 
$\nabla_m\eta_1$ would be given by the right-hand side of (\ref{ke}) plus additional terms. It can then be seen,  
by similar manipulations as above, that these  
additional terms would violate (some of) the equations in (\ref{intsb2}).

The Sasaki-Einstein metric associated with the $SU(2)$ structure (\ref{intsb6},\ref{intsb2}) can locally 
 be put in the canonical form:
\eq{\label{sem}
ds_{SE}^2=ds^2_{KE}+u\otimes u~,
}
where $ds^2_{KE}$ is a K\"{a}hler-Einstein four-dimensional base over which $u$ is fibered. 
The connection field strength of this local $U(1)$ fibration is the K\"{a}hler form of the base, 
and is equal to $du=-2\gamma$. If in addition the orbits of the vector\footnote{ This 
is known as the Reeb vector; as follows from (\ref{ke},\ref{aa1}), it is Killing and has unit norm.}  dual to $u$ 
close and the associated $U(1)$ action is free, (\ref{sem}) extends globally and the base is a four-dimensional 
K\"{a}hler-Einstein manifold of positive curvature.

The $SU(2)$ structure (\ref{intsb6}) possesses an $SU(2)$ invariance which also 
leaves the associated metric (\ref{sem}) invariant. In order to see this, 
let us define the triplet of real three forms:
\eq{
\vec{J}:=-\frac{i}{2}\vec{\sigma}_{ij}(\eta^\dagger_i\Gamma_{(2)}\eta_{j})
~,}
where $\vec{\sigma}$ is a triplet of Pauli matrices, so that $\alpha=J^{(2)}$, $\beta=J^{(1)}$, $\gamma=J^{(3)}$. Up to normalization these obey:
\eq{\label{crel}
J^{(a)}\wedge J^{(b)}=\delta^{ab}\mathrm{vol}_4~,~~~a,b=1,2,3
~,}
where vol$_4$ is the volume element of the four-dimensional base of the fibration (\ref{sem}). It is a straightforward computation 
to show that under infinitesimal $SU(2)$ transformations of the spinors $\eta_{1,2}$,
\eq{\label{sdoublet}
\delta\eta_{i}=\frac{i}{2}\delta\vec{\theta}\cdot\vec{\sigma}_{ij}\eta_j
~,}
the forms $J^{(a)}$ transform as a vector of 
$SO(3)$:
\eq{\label{jvec}
\delta \vec{J}=-\delta\vec{\theta}\times\vec{J}
~.}
Both transformations (\ref{sdoublet},\ref{jvec}) leave the associated metric invariant.

One can also see the $SU(2)$ invariance of the metric directly as follows: Choosing the 
orthonormal frame so that:
\eq{
J^{(1)}+iJ^{(2)}=e_1\wedge e_2~;~~~~~J^{(3)}=-\frac{i}{2}\left(e_1\wedge e^*_1+
e_2\wedge e^*_2
\right)
~,}
the metric (\ref{sem}) can be written as:
\eq{\label{semb}
ds^2_{SE}=e_1\otimes e^*_1+
e_2\otimes e^*_2+u\otimes u~.
}
It is then straightforward to read off the action of  (\ref{jvec}) 
on the $e_i$'s and show that it leaves the metric invariant. For example, 
infinitesimal rotations of the form (\ref{jvec})  in the $(1,2)$ plane 
imply $\delta e_{1,2}=\frac{i}{2}\delta\theta~\! e_{1,2}$, and similarly 
for the $(1,3)$ and $(2,3)$ planes.


We can now state  the precise relation between the metric   of $\mathcal{M}_6$ 
associated with the triplet $( K,\widetilde{J},\omega)$ in (\ref{sb1}) and the  metric associated with 
the Sasaki-Einstein structure (\ref{intsb6}). From the discussion above 
and the fact that triplet 
$\mathrm{Re}\omega$, $\mathrm{Im}\omega$, $\widetilde{J}$ 
is obtained up to rescalings from the triplet $\alpha$, $\beta$, $\gamma$ by an $SO(3)$ rotation, 
it follows that the metric on $\mathcal{M}_6$ is given by:
\eq{
ds^2(\mathcal{M}_6)=e^{2A}\left(3ds^2_{KE}+\xi^2(t)~\!u\otimes u\right)+e^{2A}~\!\d t\otimes \d t
~,}
where $ds^2_{KE}$ is the four dimensional K\"{a}hler-Einstein metric of eq.~(\ref{sem}). 
This is a codimension-one foliation with five-dimensional leaves 
$\mathcal{M}_5$. As advertised, 
the metric $ds^2(\mathcal{M}_5)$ is obtained from the Sasaki-Einstein metric (\ref{sem}) 
by a warping given by $e^{2A(t)}$ and a squashing given by $\xi(t)/\sqrt{3}$. 
The metrics in (\ref{sb5def}), (\ref{s5}) are obtained by the same reasoning.

\section{Massless IIA with strict $SU(3)$ structure}\label{iiaside}

For non-zero Romans mass, the case of strict 
$SU(3)$ structure was considered in \cite{lt}. As was shown in that reference, 
the dilaton and warp factor must then be constant. 
In the case of zero Romans mass it is possible 
to generalize this to include non-constant warp factor and dilaton,  provided:
\eq{\label{a1}
\phi=3A~,
}
as can be seen from e.g. eq.~(2.16) of \cite{ltb}. Moreover, in the conventions of \cite{ltb} which 
we follow here, taking the zero Romans mass limit requires setting the real part of the 
inverse $\ads$ radius  to zero:
\eq{
W=i\mathrm{Im}W
~.}
The only non-zero fluxes are given by:
\eq{\spl{\label{nzf}
e^\phi F_2&=-2\d A\lrcorner \mathrm{Re}\Omega-\frac{1}{3}e^{-A}\mathrm{Im}WJ+e^\phi\widetilde{F}\\
e^\phi F_6&=-3e^{-A}\mathrm{Im}W vol_6
~,}}
where $\widetilde{F}_2$ is a primitive piece. The non-zero $SU(3)$ torsion classes are given by:
\eq{\spl{
\mathcal{W}_1&=-\frac{4i}{3}e^{-A}\mathrm{Im}W\\
\mathcal{W}_2&=ie^\phi \widetilde{F}\\
\mathcal{W}_5&=(\d A)^{1,0}
~,}}
so that:
\eq{\spl{\label{torsiia}
\d J&=2e^{-A}\mathrm{Im}W\mathrm{Re}\Omega\\
\d\left(e^{-A}\mathrm{Im}\Omega\right)&=-\frac{4}{3}e^{-2A}\mathrm{Im}WJ\wedge J+e^{2A} \widetilde{F}\wedge J\\
\d\left(e^{-A}\mathrm{Re}\Omega\right)&=0
~.}}
Requiring $\d^2=0$ on $(J,\Omega)$ is equivalent to:
\eq{
J\wedge \d (e^{2A}\widetilde{F}-\frac{4}{3}e^{-2A}\mathrm{Im}W J)=0
~.}
The Bianchi identities for the RR six-form:
\eq{
\d F_6=\d\left(e^{4A}\star_6 F_6\right)=0
~,}
can be seen to be automatically satisfied. Moreover, the Bianchi identity for the RR two-form:
\eq{
\d\left(e^{4A}\star_6 F_2\right)
~,}
is also automatically satisfied, as can be seen by taking (\ref{torsiia}) into account and using:
\eq{
\star_6 \left( \d A\lrcorner \mathrm{Re}\Omega\right)=-\d A\wedge\mathrm{Im}\Omega
~.}
The identity above can  be derived by expressing $\Omega$ in terms of the local $SU(2)$ structure,   
$\Omega=-i\omega\wedge K$, where $(\d A)^{1,0}=\frac{1}{2}K(K^*\partial A)$. The remaining Bianchi identity:
\eq{\label{bi2}
\d F_2=0~,
}
does not follow automatically and therefore imposes an additional constraint.

\section{Spinor conventions in five dimensions}\label{spinorconventions}

In this section we list our spinor conventions in five Euclidean dimensions, which 
are used in appendix \ref{se}. 

The irreducible spinor representation is four-dimensional pseudoreal.  
The charge conjugation and the gamma matrices obey:
\eq{
C^{Tr}=-C~;~~~~~(\Gamma_mC)^{Tr}=-\Gamma_mC
~.}
The five-dimensional Hodge operator acts on the gamma matrices as follows:
\eq{
\star\Gamma^{(5-k)}=(-)^{\frac{1}{2}k(k-1)}\Gamma^{(k)}
~,}
where $\Gamma^{(k)}$ is the antisymmetrized product of $k$ gamma matrices.

The Fierz identity reads:
\eq{\label{fier}
\chi\widetilde{\psi}=-\frac{1}{4}\left\{
(\widetilde{\psi}\chi)+(\widetilde{\psi}\Gamma_m\chi)\Gamma^m+\frac{1}{2}(\widetilde{\psi}\Gamma_{mn}\chi)\Gamma^{mn}
\right\}
~,}
for any pair of commuting spinors $\chi$, $\psi$, where we have defined:
\eq{\label{wtdef}
\widetilde{\psi}:=\psi^{Tr}C^{-1}
~.}

\section{Useful identities}\label{app}

The following identities are useful in verifying the Bianchi identities 
of the various solutions presented in the main text:
\eq{\spl{
\star (\wj\wedge\wj\wedge K)&=-2i K\\
\star(\omega\wedge K)&=-i\omega\wedge K\\
\star(\omega^*\wedge K)&=-i\omega^*\wedge K\\
\star(\wj\wedge K)&=-i\wj\wedge K\\
\star J&=-\frac{1}{2}J\wedge J\\
\star\omega&=-\frac{i}{2}\omega\wedge K\wedge K^*\\
\star K&=-\frac{i}{2}\wedge\wj\wedge\wj\wedge K
~,}}
where the Hodge star above is with respect to the internal 
six-dimensional metric.

\end{document}